\documentstyle[aps,epsfig,amsfonts,epsf,feynmp]{revtex}

\newcommand{\nc}{\newcommand}
\newcommand{\scs}{\scriptstyle}

\nc{\la}{\label} \nc{\r}{\ref} \nc{\no}{\nonumber} \nc{\ci}{\cite}
\nc{\setval}{\fmfset{wiggly_len}{1.5mm}
\fmfset{arrow_len}{1.5mm}\fmfset{arrow_ang}{13}
\fmfset{dash_len}{1.5mm}\fmfpen{0.125mm}\fmfset{dot_size}{1thick}
}
\newcommand{\ddfermi}[1]{\frac{\delta^2 #1}{\delta
\parbox{10mm}{\centerline{
\begin{fmfgraph*}(5,3)
\setval \fmfleft{v1} \fmfright{v2} \fmf{fermion}{v2,v1}
\fmfv{decor.size=0,label={\footnotesize 1},l.dist=0.5mm}{v1}
\fmfv{decor.size=0,label={\footnotesize 2},l.dist=0.5mm}{v2}
\end{fmfgraph*}
}}\,\delta
\parbox{10mm}{\centerline{
\begin{fmfgraph*}(5,3)
\setval \fmfleft{v1} \fmfright{v2} \fmf{fermion}{v2,v1}
\fmfv{decor.size=0,label={\footnotesize 3},l.dist=0.5mm}{v1}
\fmfv{decor.size=0,label={\footnotesize 4},l.dist=0.5mm}{v2}
\end{fmfgraph*}
}} }}
\newcommand{\ddbfermi}[1]{\frac{\delta^2 #1}{\delta
\parbox{10mm}{\centerline{
\begin{fmfgraph*}(5,3)
\setval \fmfleft{v1} \fmfright{v2} \fmf{fermion}{v2,v1}
\fmfv{decor.size=0,label={\footnotesize 1},l.dist=0.5mm}{v1}
\fmfv{decor.size=0,label={\footnotesize 2},l.dist=0.5mm}{v2}
\end{fmfgraph*}
}}\,\delta
\parbox{10mm}{\centerline{
\begin{fmfgraph*}(5,3)
\setval \fmfleft{v1} \fmfright{v2} \fmf{boson}{v2,v1}
\fmfv{decor.size=0,label={\footnotesize $\bar{1}$},l.dist=0.5mm}{v1}
\fmfv{decor.size=0,label={\footnotesize $\bar{2}$},l.dist=0.5mm}{v2}
\end{fmfgraph*}
}} }}
\nc{\dvertex}[4]{\frac{\delta #1}{\rule[0pt]{0pt}{15pt}\delta
\parbox{10mm}{\centerline{
\begin{fmfgraph*}(4,3.464)
\setval \fmfforce{1w,0h}{v1} \fmfforce{0w,0h}{v2}
\fmfforce{0.5w,1h}{v3} \fmfforce{0.5w,0.2886h}{vm}
\fmf{fermion}{v1,vm} \fmf{fermion}{vm,v2} \fmf{photon}{v3,vm}
\fmfv{decor.size=0,label={\footnotesize #3},l.dist=0.5mm}{v1}
\fmfv{decor.size=0,label={\footnotesize #4},l.dist=0.5mm}{v2}
\fmfv{decor.size=0,label={\footnotesize #2},l.dist=0.5mm}{v3}
\fmfdot{vm}
\end{fmfgraph*}}}\rule[0pt]{0pt}{15pt}}}
\nc{\dbphi}[3]{\frac{\delta #1}{\delta
\parbox{10mm}{\centerline{
\begin{fmfgraph*}(5,3)
\setval \fmfleft{v1} \fmfright{v2} \fmf{photon}{v2,v1}
\fmfv{decor.size=0,label={\footnotesize #2},l.dist=0.5mm}{v1}
\fmfv{decor.size=0,label={\footnotesize #3},l.dist=0.5mm}{v2}
\end{fmfgraph*}
}}}}
\nc{\ddvertex}[6]{\frac{\delta^2 #1}{\rule[0pt]{0pt}{15pt} \delta
\parbox{10mm}{\centerline{
\begin{fmfgraph*}(5,3)
\setval \fmfleft{v1} \fmfright{v2} \fmf{fermion}{v2,v1}
\fmfv{decor.size=0,label={\footnotesize #2},l.dist=0.5mm}{v1}
\fmfv{decor.size=0,label={\footnotesize #3},l.dist=0.5mm}{v2}
\end{fmfgraph*}}}
\,\,\, \delta \parbox{10mm}{\centerline{
\begin{fmfgraph*}(4,3.464)
\setval \fmfforce{1w,0h}{v1} \fmfforce{0w,0h}{v2}
\fmfforce{0.5w,1h}{v3} \fmfforce{0.5w,0.2886h}{vm}
\fmf{fermion}{v1,vm} \fmf{fermion}{vm,v2} \fmf{boson}{v3,vm}
\fmfv{decor.size=0,label={\footnotesize #6},l.dist=0.5mm}{v1}
\fmfv{decor.size=0,label={\footnotesize #5},l.dist=0.5mm}{v2}
\fmfv{decor.size=0,label={\footnotesize #4},l.dist=0.5mm}{v3}
\fmfdot{vm}
\end{fmfgraph*}}}
}}
\nc{\dephi}[3]{\frac{\delta #1}{\delta
\parbox{10mm}{\centerline{
\begin{fmfgraph*}(5,3)
\setval \fmfleft{v1} \fmfright{v2} \fmf{electron}{v2,v1}
\fmfv{decor.size=0,label={\footnotesize #2},l.dist=0.5mm}{v1}
\fmfv{decor.size=0,label={\footnotesize #3},l.dist=0.5mm}{v2}
\end{fmfgraph*}
}}}}
\nc{\cdphi}[2]{\frac{\delta #1}{\delta
\parbox{10mm}{\centerline{
\begin{fmfgraph*}(5,3)
\setval \fmfleft{v1} \fmfright{v2} \fmf{photon}{v2,v1}
\fmfv{decor.size=0,label={\footnotesize #2},l.dist=0.5mm}{v2}
\fmfv{decor.shape=cross,decor.filled=shaded,decor.size=3thick}{v1}
\end{fmfgraph*}
}}}}
\newcommand{\dfermi}[3]{\frac{\delta #1}{\delta
\parbox{10mm}{\centerline{
\begin{fmfgraph*}(5,3)
\setval \fmfleft{v1} \fmfright{v2} \fmf{fermion}{v2,v1}
\fmfv{decor.size=0,label={\footnotesize #2},l.dist=0.5mm}{v1}
\fmfv{decor.size=0,label={\footnotesize #3},l.dist=0.5mm}{v2}
\end{fmfgraph*}
}}}}

\newcommand{\gl}{Ginzburg--Landau }
\newcommand{\GL}{GINZBURG--LANDAU }
\newcommand{\g}{g}
\newcommand{\e}{e}
\newcommand{\esqr}{e^2}
\newcommand{\elm}{electromagnetic }
\newcommand{\f}{\frac}
\newcommand{\p}{\delta}
\newcommand{\be}{\begin{equation}}
\newcommand{\ee}{\end{equation}}
\newcommand{\beqn}{\begin{eqnarray}}
\newcommand{\eeqn}{\end{eqnarray}}
\newcommand{\bcross}{\phi^{\dag}}
\newcommand{\boson}{\phi^{\vphantom{\dag}}}
\newcommand{\Db}{{\cal D}\boson}
\newcommand{\Dbcross}{{\cal D}\bcross}
\newcommand{\DA}{{\cal D}A}
\newcommand{\fullint}{\int\!\!\Db\Dbcross\DA}
\newcommand{\Amu}{A_{\mu}}
\newcommand{\A}[1]{A_{\bar{#1}}}
\newcommand{\invG}[1]{G^{-1}_{#1}}
\newcommand{\G}[1]{G_{#1}}
\newcommand{\invD}[1]{D^{-1}_{\overline{#1}}}
\newcommand{\invDnoindex}{D^{-1}}
\newcommand{\D}[1]{D_{\overline{#1}}}
\newcommand{\Vg}[1]{V_{#1}}
\newcommand{\Ve}[2]{H_{\bar{#1}{#2}}}
\newcommand{\Vesqr}[2]{F_{\overline{#1}{#2}}}
\newcommand{\Vgnum}{V_g}
\newcommand{\Venum}{V_e}
\newcommand{\Vesqrnum}{V_{e^2}}
\newcommand{\quarknum}{I_{\rm q}}
\newcommand{\photonnum}{I_{\gamma}}
\newcommand{\loopnum}{L}
\newcommand{\deltaA}[1]{\delta_{\overline{#1}}}
\newcommand{\Z}{Z}
\newcommand{\W}{W}
\newcommand{\Wfree}{\W_0}
\newcommand{\Wint}{\W_{\rm int}}

\newcommand{\Tr}{\rm Tr}
\newcommand{\onehalf}{\frac{1}{2}}
\newcommand{\WL}[1]{\W^{(#1)}}
\newcommand{\WLd}[2]{\WL{#1,#2}}
\newcommand{\ep}{\epsilon}

\begin{document}

\everymath={\displaystyle}

\title{Recursive Graphical Construction for Feynman Diagrams and
Their Weights in Ginzburg-Landau Theory} 

\author{H. Kleinert \and A. Pelster}
\address{Institut f\"ur Theoretische Physik, Arnimallee 14 D-14195 Berlin, Germany\\
E-mails: kleinert@physik.fu-berlin.de, pelster@physik.fu-berlin.de}
\author{B. Van den Bossche}
\address{Physique Nucl\'eaire Th\'eorique, B5,
Universit\'e de Li\`ege Sart-Tilman,
 4000 Li\`ege, Belgium\\ E-mail: bvandenbossche@ulg.ac.be\\
and\\
Institut f\"ur Theoretische Physik, Arnimallee 14 D-14195 Berlin, Germany\\
E-mail: bossche@physik.fu-berlin.de}

\maketitle

\begin{abstract}
The free energy of the \gl theory satisfies a
nonlinear functional differential equation which is turned
into a recursion relation. The latter is solved graphically
order by order in the loop expansion to find all connected vacuum
diagrams, and their corresponding weights. In this way
we determine the connected vacuum diagrams and their weights up to four loops. 
\end{abstract}

\date{\today}

\section{Introduction}
\label{introduction}
Recently, two of us (H.K. and B.V.d.B.) have determined the two-loop
effective potential of $O(N)$-symmetric scalar quantum electrodynamics in $4-\ep$ dimensions
and its $\ep$-expansion \cite{kvdbeffpot}.
Going to higher loop orders requires the calculation of many complicated
Feynman integrals associated with vacuum diagrams. 
In this paper, we show how to find all these diagrams and their weights with the help of
a simple recursive technique
described in detail in Refs.~\cite{sv95,phi4,qed,bkphi4,asym}.
The basics of the method were formulated in Refs.~\cite{klrecrel,russianbook}. For simplicity, 
the present work is restricted to the theory of a single complex scalar field coupled
to electromagnetism, known as Ginzburg-Landau theory or as scalar quantum electrodynamics. 
In $D=3$ dimensions, this describes the
physics of superconductors. Without electromagnetism, we recover the physics of superfluid helium.
Here we shall restrict ourselves to the
symmetric phase, where the $O(N)$-symmetry is unbroken, which describes the system
above the critical temperature $T_c$. The more general situation will be dealt with elsewhere.

For a complex boson field, propagators are represented by
oriented lines, as in ordinary quantum electrodynamics \cite{qed}. When comparing the two expansions we must,
however, drop a minus sign for each fermion loop. In addition, the \gl theory contains four-point functions of the
scalar field which are in principle treated in $\phi^4$-theory \cite{phi4}. However,
since we deal here with complex fields, the weight of
the corresponding graphs cannot be taken from Ref.~\cite{phi4}. There is furthermore a completely
new vertex in scalar QED: the quartic seagull coupling between two photon and two
scalar fields. By a replacement of the photon lines by scalar lines, they become
equivalent to the diagrams coming from the $\phi^4$-vertex, apart, of course,
from different weights.

The paper is organized as follows. In Section~\ref{gltheory} we define more precisely the theory to be studied.
In Section~\ref{BASIC} we introduce basic functional derivatives which allow
in Section~\ref{generatingequation} to derive a functional differential equation for the free energy. From this follows
a recursion relation which allows to find all vacuum diagrams of a given order $L$ from
those of the previous orders. This equation is formulated graphically in 
Section~\ref{generatinggraphs}, where we also determine the vacuum
diagrams and the corresponding weights up to  four loops.

\section{\protect \GL theory}
\label{gltheory}
The generalized \gl theory deals with a self-interacting
complex scalar field coupled minimally to an
electromagnetic vector potential.
The theory contains two coupling constants $\g$ and $\e$. 
The field expectation value $\langle\phi\rangle$ vanishes above $T_c$
and the theory has  three types of vertices.

The physics can be extracted from the partition function
\be \Z=\int\Db\Dbcross\DA\exp\left( -E \right),
\label{partitionfn}
\ee
with thermal fluctuations being governed by the energy functional
\be
E(\bcross,\boson,\Amu)=\bcross_1\invG{12}\boson_2+\frac{\g}{4}
\Vg{1234}\bcross_1\boson_2\bcross_3\boson_4
+\onehalf\A{1}\invD{12}\A{2}+\e\Ve{1}{12}\A{1}\bcross_1\boson_2
+\frac{\esqr}{2} \Vesqr{12}{12}\A{1}\A{2}\bcross_1\boson_2.
\label{energy}
\ee
Here and below, overlined indices are used for the \elm field. We
work in a covariant gauge and assume the photon propagator to have the appropriate form (see Ref.~\cite{qed}).
Using the same notation as in Refs.~\cite{phi4,qed}, 
we keep all equations as compact as possible by assuming
Einstein's summation convention not only for the internal
degrees of freedom, but also for the space-time indices, for which repeated indices imply
an overall integral sign. In this notation the functional matrices in the energy functional (\ref{energy}) 
have the following symmetries:

\beqn
\invD{12}&=&\invD{21},\label{invDsym}\\
\Vg{1234}&=&\Vg{3214}=\Vg{1432}=\Vg{3412},\label{Vgsym}\\
\Vesqr{12}{12}&=&\Vesqr{21}{12}.\label{Vesqrsym}
\label{vertices}
\eeqn

\section{Basic Functional Derivatives}\label{BASIC}

The recursion relation will be derived by performing functional differentiations of the partition function with
respect to the propagators and their inverse as well as the vertices. 
The basic properties of these derivatives are shown in the following
equations:

\subsection*{Scalar sector}
\label{scalarsector}
\beqn
\f{\p\invG{12}}{\p\invG{34}}&=&\delta_{13}\delta_{24},\label{invGtoGfirst}\\
\f{\p\G{12}}{\p\G{34}}&=&\delta_{13}\delta_{24},\\
\f{\p\G{12}}{\p\invG{34}}&=&-\G{13}\G{42},\\
\f{\p}{\p\invG{34}}&=&-\G{13}\G{42}\f{\p}{\p\G{12}}.\label{invGtoG}
\eeqn

\subsection*{Photon sector}
\label{photonsector}
\beqn
\f{\p\invD{12}}{\p\invD{34}}&=&\onehalf\left(\deltaA{13}\deltaA{42}+
\deltaA{14}\deltaA{32}\right),\\
\f{\p\D{12}}{\p\D{34}}&=&\onehalf\left(\deltaA{13}\deltaA{42}+
\deltaA{14}\deltaA{32}\right),\\
\f{\p\D{12}}{\p\invD{34}}&=&-\onehalf\left(\D{13}\D{42}+
\D{14}\D{32}\right),\\
\f{\p}{\p\invD{34}}&=&-\D{13}\D{42}\f{\p}{\p\D{12}}.\label{invDtoD}
\eeqn

\subsection*{Vertex derivatives}

\begin{eqnarray}
\f{\p\Ve{1}{12}}{\p \Ve{2}{34}}&=&\deltaA{12}\delta_{13}\delta_{24}, \\
\frac{\delta V_{1234}}{\delta V_{5678}} &=& \frac{1}{4} \left( 
\delta_{15} \delta_{26} \delta_{37} \delta_{48} + 
\delta_{17} \delta_{26} \delta_{35} \delta_{48} + 
\delta_{15} \delta_{28} \delta_{37} \delta_{46} + 
\delta_{17} \delta_{28} \delta_{35} \delta_{46} \right),\\
\frac{\delta F_{\overline{12}12}}{\delta F_{\overline{34}34}} &=& \frac{1}{2} \left( 
\delta_{\overline{13}} \delta_{\overline{24}} \delta_{13} \delta_{24} + 
\delta_{\overline{14}} \delta_{\overline{23}} \delta_{13} \delta_{24} \right).\\ && \nonumber
\end{eqnarray}

This is a direct extension of relations used in Refs.~\cite{phi4,qed,bkphi4,asym}.
The main difference comes from the fact that the previous scalar propagators were symmetric in the indices
whereas the present propagators describing complex fields are not. By the chain rule of differentiations,
we can then find the derivative of any functional with respect to the propagators and their inverse as
well as the vertices.

\section{Functional differential equation for free energy}
\label{generatingequation}

With the definitions given in the previous section, we are now prepared to
derive the graphical recursion relation for the vacuum graphs. We start from the identity
\be
\fullint \f{\p}{\p\bcross_1}\left[
\bcross_2\exp\left(
-E
\right)
\right] = 0,
\label{scalarSDdef1}
\ee
and obtain
\be
\fullint\left(
\delta_{12}-\bcross_2\f{\p E}{\p\bcross_1}
\right)\exp\left(
-E
\right)= 0
\label{scalarSDdef2}
\ee
which, using the \gl energy functional (\ref{energy}), leads
to the linear equation
\be
\fullint\left(
\delta_{12}-\bcross_2\invG{13}\boson_3
- \frac{\g}{2} \Vg{1345}\bcross_2\boson_3\bcross_4\boson_5
- \frac{\esqr}{2} \Vesqr{12}{13}\A{1}\A{2}\bcross_2\boson_3
-\e\Ve{1}{13}\A{1}\bcross_2\boson_3
\right)\exp\left(-E\right)=0.
\label{scalarSDdef3}
\ee
Rewritten in terms of functional derivatives with respect to the inverse of the propagators,
this becomes
\be
\fullint\left(
\delta_{12}+\invG{13}\f{\p}{\p\invG{23}}
- \frac{\g}{2} \Vg{1345}\f{\p}{\p\invG{23}}\f{\p}{\p\invG{45}}
- \esqr \Vesqr{12}{13}\f{\p}{\p\invD{12}}\f{\p}{\p\invG{23}}
+\e\Ve{1}{13}\A{1}\f{\p}{\p\invG{23}}
\right)\exp\left(
-E
\right)=0.
\label{scalarSDdef4}
\ee
The last part of the above equation may be replaced by a three-vertex
derivative. However,
proceeding in this way would not lead to an iterative generation of
diagrams. For this, it is
necessary to consider a second identity:
%
\be
\fullint \f{\p}{\p\A{1}}\exp\left(
-E\right)= 0,
\label{photonSDdef1}
\ee
from which  we obtain
\be
\fullint
\left(
\invD{12}\A{2}+\e\Ve{1}{12}\bcross_1\boson_2
+\esqr\Vesqr{12}{12}\A{2}\bcross_1\boson_2\right)\exp\left(
-E
\right)=0,
\label{photonSDdef2}
\ee
or, equivalently,
\be
\fullint
\left(
\invD{12}\A{2}-\e\Ve{1}{12}\f{\p}{\p\invG{12}}
-\e\Vesqr{12}{12} \f{\p}{\delta \Ve{2}{12}}
\right)\exp\left(-E
\right)=0.
\label{photonSDdef3}
\ee
Inserting the photon-field expectation value~(\ref{photonSDdef3}) 
into~(\ref{scalarSDdef4}), we obtain
\beqn
&&\Bigg(
\delta_{12}+\invG{13}\f{\p}{\p\invG{23}}
-\frac{\g}{2}\Vg{1345}\f{\p}{\p\invG{23}}\f{\p}{\p\invG{45}}
-\esqr\Vesqr{12}{13}\f{\p}{\p\invD{12}}\f{\p}{\p\invG{23}}
\nonumber\\
&&\hspace{2cm}\mbox{}+
\esqr\Ve{1}{13}\D{12}\Ve{2}{45}\f{\p}{\delta \invG{23}}\f{\p}{\delta \invG{45}}
+\esqr\Ve{1}{13}\D{12}\Vesqr{23}{45}\f{\p}{\delta \invG{23}}\f{\p}{\delta \Ve{3}{45}}
\Bigg)\Z=0.
\label{Zdiffequation}
\eeqn
This equation is linear, but it leads to a huge number of diagrams,
both connected and disconnected.
We remove the disconnected ones by introducing 
the free energy $W$ as generating functional of the connected Green functions as
\be
\Z\equiv\exp\left(
\W
\right), \qquad \W=\Wfree+\Wint,
\label{Wdefinition}
\ee
where $\Wfree$ is the free field part
\be
\Wfree=-\Tr\ln\invG{}-\onehalf\Tr\ln\invDnoindex,
\ee
with \Tr\ being a shorthand notation
for the  functional trace. Working with the series
representation of the logarithm, we obtain directly the relations
\beqn
\f{\p\Wfree}{\p\invG{12}}&=&-\G{21},\label{freeinvGderiv}\\
\f{\p\Wfree}{\p\invD{12}}&=&-\onehalf\D{21},\label{freeinvDderiv} \, ,\\
\f{\p\Wfree}{\delta \Ve{1}{23}} & = & 0 \, , \label{freeinvH}
\eeqn
where the index ordering is important for the complex scalar
fields.


Introducing the decomposition~(\ref{Wdefinition}) in~(\ref{Zdiffequation}) and using the
relations (\ref{freeinvGderiv})--(\ref{freeinvH}),
we obtain a nonlinear functional differential equation for the interacting part $\Wint$:
\beqn
&&\invG{13}\f{\p\Wint}{\p\invG{23}}-\frac{\g}{2}\Vg{1345}\left[
2\G{34}\G{52}-\left(
\G{32}\f{\p\Wint}{\p\invG{45}}+\G{54}\f{\p\Wint}{\p\invG{23}}
\right)
+\f{\p^2\Wint}{\p\invG{23}\p\invG{45}}
+\f{\p\Wint}{\p\invG{23}}\f{\p\Wint}{\p\invG{45}}
\right]\nonumber\\
&&\hspace{0.5cm}\mbox{}-\esqr\Vesqr{12}{13}
\left[
\onehalf\D{21}\G{32}-\left(
\onehalf\D{21}\f{\p\Wint}{\p\invG{23}}
+\G{32}\f{\p\Wint}{\p\invD{12}}
\right)
+\f{\p^2\Wint}{\p\invD{12}\p\invG{23}}
+\f{\p\Wint}{\p\invD{12}}\f{\p\Wint}{\p\invG{23}}
\right]\nonumber\\
&&\hspace{0.5cm}\mbox{}+\esqr\Ve{1}{13}\D{12}\Ve{2}{45}
\left[
\G{52}\G{34}+\G{32}\G{54}-\left(
\G{32}\f{\p\Wint}{\p\invG{45}}+\G{54}\f{\p\Wint}{\p\invG{23}}
\right)
+\f{\p^2\Wint}{\p\invG{23}\p\invG{45}}
+\f{\p\Wint}{\p\invG{23}}\f{\p\Wint}{\p\invG{45}}
\right]\nonumber\\
&&\hspace{0.5cm}\mbox{}+\esqr\Ve{1}{13}\D{12}\Vesqr{23}{45}
\left[
-\G{32}\f{\p\Wint}{\p\Ve{3}{45}}
+\f{\p^2\Wint}{\p\invG{23}\p\Ve{3}{45}}
+\f{\p\Wint}{\p\invG{23}}\f{\p\Wint}{\p\Ve{3}{45}}
\right]=0.
\eeqn
With the help of Eqs.~(\ref{invGtoG}) and~(\ref{invDtoD}), 
this equation becomes
\beqn
&&-\G{12}\f{\p\Wint}{\p\G{12}}-\frac{\g}{2}\Vg{1234}\left[
2\G{23}\G{41}+4\G{21}\G{53}\G{46}\f{\p\Wint}{\p\G{56}}
+\G{51}\G{26}\G{73}\G{48}
\left(
\f{\p^2\Wint}{\p\G{56}\p\G{78}}
+\f{\p\Wint}{\p\G{56}}\f{\p\Wint}{\p\G{78}}
\right)
\right]\nonumber\\
&& -\esqr\Vesqr{12}{12}
\left[
\onehalf\D{21}\G{21}
+\onehalf\D{21}\G{31}\G{24}\f{\p\Wint}{\p\G{34}}
+\G{21}\D{31}\D{24}\f{\p\Wint}{\p\D{34}}
+\D{31}\D{24}\G{31}\G{24}
\left(
\f{\p^2\Wint}{\p\D{34}\p\G{34}}
+\f{\p\Wint}{\p\D{34}}\f{\p\Wint}{\p\G{34}}
\right)
\right]\nonumber\\
&&+\esqr\Ve{1}{12}\D{12}\Ve{2}{34}
\left[
\G{41}\G{23}+\G{21}\G{43}
+2\left(
\G{21}\G{53}+\G{51}\G{23}
\right)\G{46}\f{\p\Wint}{\p\G{56}}
+\G{51}\G{26}\G{73}\G{48} \left(
\f{\p^2\Wint}{\p\G{56}\p\G{78}}
\right. \right.
\nonumber\\&&
\left. \left.
+\f{\p\Wint}{\p\G{56}}\f{\p\Wint}{\p\G{78}}
\right) \right]
-\esqr\Ve{1}{12}\D{12}\Vesqr{23}{34}
\left[
\G{21}\f{\p\Wint}{\p\Ve{3}{34}}
+\G{51}\G{26}\left(
\f{\p^2\Wint}{\p\G{56}\p\Ve{3}{34}}
+\f{\p\Wint}{\p\G{56}}\f{\p\Wint}{\p\Ve{3}{34}}
\right)
\right]=0.
\label{masterequation}
\eeqn
From this functional differential equation, a graphical recursion relation can be derived. This is the subject
of the next section.

\section{Graphical recursion relation}
\label{generatinggraphs}

From~(\ref{masterequation}), we can derive a graphical recursion
relation for the
connected vacuum diagrams. When considering the
loop expansion of the interaction part $\Wint$, one
term is of one loop number larger than the other terms, as we now
show.

The operators $\G{12}\p/\p\G{12}$ and $\D{12}\p/\p\D{12}$ simply
count the number
of scalar lines $\quarknum$ and photon lines $\photonnum$, respectively, in a
given
diagram.
These numbers can be
extracted from the number and type of vertices. 
Denoting by $\Vgnum$, $\Venum$ and
$\Vesqrnum$
the number of $\g$, $\e$ and $\esqr$ vertices, we have the following 
counting rules.
The Yukawa vertex $\Ve{1}{12}$ has
one photon line and
two scalar lines, while the quartic photon-scalar vertex $\Vesqr{12}{12}$ 
has two photon lines
and two scalar
lines. Furthermore two vertex lines are necessary to produce an internal line
when combining
vertices. We have the obvious relation $2\photonnum=2\Vesqrnum+\Venum$. Taking the
quartic scalar self-interaction $V_{1234}$ into account, we have also
$2\quarknum=2\Vesqrnum+2\Venum+4\Vgnum$.
An odd number of photon fields gives no
contribution to the free energy. Thus the vertex $\e$ enters with even power.
The number of loops is then
easily found to be $\loopnum-1=\Vgnum+\Venum/2+\Vesqrnum.\label{loopnumber}$
Together, we have the counting rules
\beqn
\photonnum&=&\Vesqrnum+\f{\Venum}{2},\\
\quarknum&=&\Vesqrnum+\Venum+2\Vgnum,\\
\loopnum-1&=&\Vesqrnum+\f{\Venum}{2}+\Vgnum.
\eeqn
These relations can be inverted to give the number of each type
of vertex as a function of the number of loops, scalar, and
photon internal lines:
\beqn
\Vesqrnum&=&2(\loopnum-1)-\quarknum,\\
\f{\Venum}{2}&=&\quarknum+\photonnum-2(\loopnum-1),\\
\Vgnum&=&\loopnum-1-\photonnum.
\eeqn
It is now clear that, to count the number of scalars in a given
diagram, it is only necessary to know the loop order, as well as
the number of the quartic
$\esqr$ photon-scalar seagull vertices:
\be
\quarknum=2(\loopnum-1)-\Vesqrnum.
\label{Eqquarknumber}
\ee
The vertices $\Venum$ are not
taken into account when counting the quartic
vertices $\Vesqrnum$, which,
by definition, count the two photon-two scalar
vertices only.
The relation~(\ref{Eqquarknumber}) is interesting,
since, for complicated diagrams, with a large
loop-order, it is much less involved to count the number of quartic
$\esqr$ photon-scalar
seagull vertices than the number of scalar lines.

We are now ready to demonstrate that equation~(\ref{masterequation})
allows for a recursive solution. We form
the loop expansion
\be
\Wint=\sum_{\loopnum=2}^{\infty}\g^{\loopnum-1}\WL{\loopnum},
\label{Wintloop}
\ee
supposing that the vertices $\e$ and $\esqr$ are
of order $\sqrt{\g}$ and $\g$, respectively. This is because
the relevant parameter for the loop expansion is the  inverse
temperature $\beta=1/(k_BT)$ 
which appears as a coefficient in front of the energy. 
The inverse temperature is set equal to one in this work (see
Eq.~(\ref{partitionfn})). We could have restored it to show the
loop counting. We would have seen readily that this would have
been equivalent to the statement that
$\e$ and $\esqr$ are of order
$\sqrt{\g}$ and $\g$, respectively.
In Eq.~(\ref{Wintloop}), $\WL{\loopnum}$ is a sum over the
different diagrams of a given
order $\loopnum$:
\be
\WL{\loopnum}=\sum_d(-1)^{\Vgnum+\Vesqrnum}\WLd{\loopnum}{d},
\label{negativfactor} \ee
where $d$ distinguishes between different classes of diagrams of
the same loop order. The factor $(-1)^{\Vgnum+\Vesqrnum}$ takes care of 
minus signs in the diagrams of the perturbation expansion.
Since the number of Yukawa vertices is even, it does
not enter this prefactor: $(-1)^{\Venum}=1$. Applying the scalar
number operator on $\Wint$ gives
\be
\G{12}\f{\p\Wint}{\p\G{12}}=\sum_{\loopnum=2}^{\infty}\g^{\loopnum-1}
\sum_d(-1)^{\Vgnum+\Vesqrnum}\quarknum(L,d)\WLd{\loopnum}{d} \, ,
\label{countingquarklines}
\ee
which stresses that each class of diagrams, i.e., each topology, has
its own scalar number.

By performing the loop expansion (\ref{Wintloop}), the contributions $\WL{\loopnum}$
to the negative free energy consists of all connected vacuum diagrams constructed according to
the following Feynman rules. A straight line and a wiggly line represent the free correlation
of the scalar and the photon field, respectively:
\begin{fmffile}{graph09}
\begin{eqnarray}
\parbox{20mm}{\centerline{
\begin{fmfgraph*}(7,3)
\setval
\fmfleft{v1}
\fmfright{v2}
\fmf{fermion}{v2,v1}
\fmfv{decor.size=0, label=${\scs 1}$, l.dist=1mm, l.angle=-180}{v1}
\fmfv{decor.size=0, label=${\scs 2}$, l.dist=1mm, l.angle=0}{v2}
\end{fmfgraph*}}}  
&\equiv &\quad G_{12}, \\
\parbox{20mm}{\centerline{
\begin{fmfgraph*}(7,3)
\setval
\fmfleft{v1}
\fmfright{v2}
\fmf{boson}{v1,v2}
\fmfv{decor.size=0, label=${\scs \overline{1}}$, l.dist=1mm, l.angle=-180}{v1}
\fmfv{decor.size=0, label=${\scs \overline{2}}$, l.dist=1mm, l.angle=0}{v2}
\end{fmfgraph*}}}
&\equiv &\quad D_{\overline{12}}.
\end{eqnarray}
The vertices are correspondingly pictured by
\begin{eqnarray}
\parbox{15mm}{\centerline{
\begin{fmfgraph}(5,4.33)
\setval
\fmfforce{1w,0h}{v1}
\fmfforce{0w,0h}{v2}
\fmfforce{0.5w,1h}{v3}
\fmfforce{0.5w,0.2886h}{vm}
\fmf{fermion}{v1,vm,v2}
\fmf{boson}{v3,vm}
\fmfdot{vm}
\end{fmfgraph}
}}&\equiv& \quad - e \int_{\overline{1}23} H_{\overline{1}23}\, , \label{VR1} \\
\parbox{17mm}{\begin{center}
\begin{fmfgraph}(4,4)
\setval
\fmfstraight
\fmfforce{0w,0h}{o2}
\fmfforce{0w,1h}{i1}
\fmfforce{1w,0h}{o1}
\fmfforce{1w,1h}{i2}
\fmfforce{1/2w,1/2h}{v1}
\fmf{boson}{i1,v1}
\fmf{boson}{v1,i2}
\fmf{fermion}{o1,v1}
\fmf{fermion}{v1,o2}
\fmfdot{v1}
\end{fmfgraph}
\end{center}} 
&\equiv& \hspace*{0.1cm} - e^2 \int_{\overline{12}34} \, F_{\overline{12}34} \, , \label{VR2}\\
\parbox{17mm}{\begin{center}
\begin{fmfgraph}(4,4)
\setval
\fmfstraight
\fmfforce{0w,0h}{o2}
\fmfforce{0w,1h}{i1}
\fmfforce{1w,0h}{o1}
\fmfforce{1w,1h}{i2}
\fmfforce{1/2w,1/2h}{v1}
\fmf{fermion}{v1,i1}
\fmf{fermion}{i2,v1}
\fmf{fermion}{v1,o1}
\fmf{fermion}{o2,v1}
\fmfdot{v1}
\end{fmfgraph}
\end{center}} 
&\equiv& \hspace*{0.1cm} - g \int_{1234} \, V_{1234} \, . \label{VR3}
\end{eqnarray}
With these Feynman rules, the insertion of the loop expansion (\ref{Wintloop}) into (\ref{masterequation}) leads to
the following equation
\beqn
&&
\parbox{8mm}{\begin{center}
\begin{fmfgraph*}(2.5,5)
\setval \fmfstraight \fmfforce{1w,0h}{v1} \fmfforce{1w,1h}{v2}
\fmf{fermion,left=1}{v1,v2} \fmfv{decor.size=0, label=${\scs 2}$,
l.dist=1mm, l.angle=0}{v1} \fmfv{decor.size=0, label=${\scs 1}$,
l.dist=1mm, l.angle=0}{v2}
\end{fmfgraph*}
\end{center}}
\hspace*{0.3cm} g\dfermi{W^{(2)}}{1}{2} 
\hspace*{0.2cm} + \hspace*{0.1cm}
\parbox{8mm}{\begin{center}
\begin{fmfgraph*}(2.5,5)
\setval \fmfstraight \fmfforce{1w,0h}{v1} \fmfforce{1w,1h}{v2}
\fmf{fermion,left=1}{v1,v2} \fmfv{decor.size=0, label=${\scs 2}$,
l.dist=1mm, l.angle=0}{v1} \fmfv{decor.size=0, label=${\scs 1}$,
l.dist=1mm, l.angle=0}{v2}
\end{fmfgraph*}
\end{center}}
\hspace*{0.3cm} \sum_{L=3}^{\infty}g^{L-1}\dfermi{W^{(L)}}{1}{2}
\quad = \quad
%
%
\parbox{7mm}{\begin{center}
\begin{fmfgraph}(4,4)
\setval
\fmfforce{0w,1/2h}{v1}
\fmfforce{1w,1/2h}{v2}
\fmf{fermion,right=1}{v2,v1} 
\fmf{fermion,right=1}{v1,v2} 
\fmf{photon}{v2,v1}
\fmfdot{v2,v1}
\end{fmfgraph}\end{center}}
%
%
\hspace*{1mm} + \hspace*{1mm}
\parbox{15mm}{\begin{center}
\begin{fmfgraph}(12,4)
\setval
\fmfforce{1/6w,1h}{v1a}
\fmfforce{1/6w,0h}{v1b}
\fmfforce{1/3w,1/2h}{v2}
\fmfforce{2/3w,1/2h}{v3}
\fmfforce{5/6w,1h}{v4a}
\fmfforce{5/6w,0h}{v4b}
\fmf{fermion,right=1}{v1a,v1b} 
\fmf{plain,right=1}{v1b,v1a} 
\fmf{plain,right=1}{v1,v2}
\fmf{photon}{v2,v3}
\fmf{fermion,right=1}{v4b,v4a} 
\fmf{plain,right=1}{v4a,v4b}
\fmfdot{v2,v3}
\end{fmfgraph}\end{center}}
%
%
\hspace*{1mm} + \frac{1}{2} \hspace*{1mm} 
\parbox{11mm}{\begin{center}
\begin{fmfgraph}(8,4)
\setval
\fmfforce{1/4w,1h}{v1a}
\fmfforce{1/4w,0h}{v1b}
\fmfforce{1/2w,1/2h}{v2}
\fmfforce{1w,1/2h}{v3}
\fmf{fermion,right=1}{v1a,v1b} 
\fmf{plain,right=1}{v1b,v1a}
\fmf{photon,right=1}{v2,v3,v2}
\fmfdot{v2}
\end{fmfgraph}\end{center}}
%
%
\hspace*{1mm} + \hspace*{1mm}
\parbox{11mm}{\begin{center}
\begin{fmfgraph}(8,4)
\setval
\fmfforce{1/4w,1h}{v1a}
\fmfforce{1/4w,0h}{v1b}
\fmfforce{1/2w,1/2h}{v2}
\fmfforce{3/4w,1h}{v3a}
\fmfforce{3/4w,0h}{v3b}
\fmf{fermion,right=1}{v1a,v1b} 
\fmf{plain,right=1}{v1b,v1a}
\fmf{plain,right=1}{v3a,v3b}
\fmf{fermion,right=1}{v3b,v3a}
\fmfdot{v2}
\end{fmfgraph}\end{center}}
\quad + \quad
\no\\ &&
\quad\sum_{L=2}^{\infty}g^{L-1}\Biggl[
%
%
2 \hspace*{0.2cm}
\parbox{12mm}{\begin{center}
\begin{fmfgraph*}(10,8)
\setval \fmfstraight \fmfforce{2.5/9w,1.5/8h}{v1}
\fmfforce{2.5/9w,6.5/8h}{v2} \fmfforce{5/9w,1/2h}{v3}
\fmfforce{1w,0h}{v4} \fmfforce{1w,1h}{v5}
\fmf{fermion,left=1}{v1,v2} \fmf{plain,left=1}{v2,v1}
\fmf{fermion}{v3,v5} \fmf{fermion}{v4,v3} \fmfdot{v3}
\fmfv{decor.size=0, label=${\scs 2}$, l.dist=1mm, l.angle=0}{v4}
\fmfv{decor.size=0, label=${\scs 1}$, l.dist=1mm, l.angle=0}{v5}
\end{fmfgraph*}
\end{center}}
\hspace*{0.3cm} \dfermi{W^{(L)}}{1}{2}
%
%
\hspace*{0.2cm} + \hspace*{0.1cm} \frac{1}{2} \hspace*{0.2cm}
\parbox{8mm}{\begin{center}
\begin{fmfgraph*}(5,10)
\setval \fmfstraight \fmfforce{0w,1/2h}{v1} \fmfforce{1w,0h}{v2}
\fmfforce{1w,1/3h}{v3} \fmfforce{1w,2/3h}{v4} \fmfforce{1w,1h}{v5}
\fmf{fermion}{v2,v1} \fmf{fermion}{v1,v3} \fmf{fermion}{v4,v1}
\fmf{fermion}{v1,v5} \fmfdot{v1} \fmfv{decor.size=0, label=${\scs
1}$, l.dist=1mm, l.angle=0}{v5} \fmfv{decor.size=0, label=${\scs
2}$, l.dist=1mm, l.angle=0}{v4} \fmfv{decor.size=0, label=${\scs
3}$, l.dist=1mm, l.angle=0}{v3} \fmfv{decor.size=0, label=${\scs
4}$, l.dist=1mm, l.angle=0}{v2}
\end{fmfgraph*}
\end{center}}
\hspace*{0.2cm} \ddfermi{W^{(L)}}
%
%
\hspace*{0.2cm} + \hspace*{0.1cm} \frac{1}{2} \hspace*{0.2cm}
\parbox{12mm}{\begin{center}
\begin{fmfgraph*}(10,8)
\setval \fmfstraight \fmfforce{2.5/9w,1.5/8h}{v1}
\fmfforce{2.5/9w,6.5/8h}{v2} \fmfforce{5/9w,1/2h}{v3}
\fmfforce{1w,0h}{v4} \fmfforce{1w,1h}{v5}
\fmf{boson,left=1}{v1,v2,v1} \fmf{fermion}{v3,v5}
\fmf{fermion}{v4,v3} \fmfdot{v3} \fmfv{decor.size=0, label=${\scs
2}$, l.dist=1mm, l.angle=0}{v4} \fmfv{decor.size=0, label=${\scs
1}$, l.dist=1mm, l.angle=0}{v5}
\end{fmfgraph*}
\end{center}}
\hspace*{0.3cm} \dfermi{W^{(L)}}{1}{2}
\no \\ && \hspace*{1cm}
%
%
+ \hspace*{0.2cm} 
\parbox{12mm}{\begin{center}
\begin{fmfgraph*}(9,8)
\setval \fmfstraight \fmfforce{2.5/9w,1.5/8h}{v1}
\fmfforce{2.5/9w,6.5/8h}{v2} \fmfforce{5/9w,1/2h}{v3}
\fmfforce{1w,0h}{v4} \fmfforce{1w,1h}{v5}
\fmf{fermion,left=1}{v1,v2} \fmf{plain,left=1}{v2,v1}
\fmf{boson}{v3,v5} \fmf{boson}{v4,v3} \fmfdot{v3}
\fmfv{decor.size=0, label=${\scs \bar{2}}$, l.dist=1mm, l.angle=0}{v4}
\fmfv{decor.size=0, label=${\scs \bar{1}}$, l.dist=1mm, l.angle=0}{v5}
\end{fmfgraph*}
\end{center}}
\hspace*{0.3cm} \dbphi{W^{(L)}}{$\bar{1}$}{$\bar{2}$}
%
%
\hspace*{0.2cm} + \hspace*{0.2cm}
\parbox{12mm}{\begin{center}
\begin{fmfgraph*}(5,10)
\setval \fmfstraight \fmfforce{0w,1/2h}{v1} \fmfforce{1w,0h}{v2}
\fmfforce{1w,1/3h}{v3} \fmfforce{1w,2/3h}{v4} \fmfforce{1w,1h}{v5}
\fmf{boson}{v2,v1} \fmf{boson}{v1,v3} \fmf{fermion}{v4,v1}
\fmf{fermion}{v1,v5} \fmfdot{v1} \fmfv{decor.size=0, label=${\scs
1}$, l.dist=1mm, l.angle=0}{v5} \fmfv{decor.size=0, label=${\scs
2}$, l.dist=1mm, l.angle=0}{v4} \fmfv{decor.size=0, label=${\scs
\bar{1}}$, l.dist=1mm, l.angle=0}{v3} \fmfv{decor.size=0, label=${\scs
\bar{2}}$, l.dist=1mm, l.angle=0}{v2}
\end{fmfgraph*}
\end{center}}
\hspace*{0.2cm} \ddbfermi{W^{(L)}}
%
%
\hspace*{0.2cm} + \hspace*{0.1cm} 2 \hspace*{0.2cm}
\parbox{17mm}{\begin{center}
\begin{fmfgraph*}(14,8)
\setval \fmfstraight \fmfforce{2.5/14w,1.5/8h}{v1}
\fmfforce{2.5/14w,6.5/8h}{v2} \fmfforce{5/14w,1/2h}{v3}
\fmfforce{10/14w,1/2h}{v4} \fmfforce{1w,0h}{v5}
\fmfforce{1w,1h}{v6} \fmf{fermion,right=1}{v2,v1}
\fmf{plain,left=1}{v2,v1} \fmf{boson}{v3,v4} \fmf{fermion}{v5,v4}
\fmf{fermion}{v4,v6} \fmfdot{v3,v4} \fmfv{decor.size=0,
label=${\scs 1}$, l.dist=1mm, l.angle=0}{v6} \fmfv{decor.size=0,
label=${\scs 2}$, l.dist=1mm, l.angle=0}{v5}
\end{fmfgraph*}
\end{center}}
\hspace*{0.2cm} \dfermi{W^{(L)}}{1}{2}
%
%
\hspace*{0.2cm} + \hspace*{0.1cm} 2 \hspace*{0.2cm}
\parbox{10.5mm}{\begin{center}
\begin{fmfgraph*}(7.5,5)
\setval \fmfstraight \fmfforce{1/3w,0h}{v1} \fmfforce{1/3w,1h}{v2}
\fmfforce{1w,0h}{v3} \fmfforce{1w,1h}{v4}
\fmf{fermion,left=1}{v1,v2} \fmf{fermion}{v2,v4}
\fmf{fermion}{v3,v1} \fmf{boson}{v1,v2} \fmfdot{v1,v2}
\fmfv{decor.size=0, label=${\scs 1}$, l.dist=1mm, l.angle=0}{v4}
\fmfv{decor.size=0, label=${\scs 2}$, l.dist=1mm, l.angle=0}{v3}
\end{fmfgraph*}
\end{center}}
\hspace*{0.2cm} \dfermi{W^{(L)}}{1}{2}
\no \\ && \hspace*{1cm}
%
%
+ \hspace*{0.2cm}
\parbox{8mm}{\begin{center}
\begin{fmfgraph*}(5,10)
\setval \fmfstraight \fmfforce{0w,1/6h}{v1a}
\fmfforce{0w,5/6h}{v1b} \fmfforce{1w,0h}{v2}
\fmfforce{1w,1/3h}{v3} \fmfforce{1w,2/3h}{v4} \fmfforce{1w,1h}{v5}
\fmf{fermion}{v2,v1a} \fmf{fermion}{v1a,v3} \fmf{fermion}{v4,v1b}
\fmf{fermion}{v1b,v5} \fmf{boson}{v1a,v1b} \fmfdot{v1a,v1b}
\fmfv{decor.size=0, label=${\scs 1}$, l.dist=1mm, l.angle=0}{v5}
\fmfv{decor.size=0, label=${\scs 2}$, l.dist=1mm, l.angle=0}{v4}
\fmfv{decor.size=0, label=${\scs 3}$, l.dist=1mm, l.angle=0}{v3}
\fmfv{decor.size=0, label=${\scs 4}$, l.dist=1mm, l.angle=0}{v2}
\end{fmfgraph*}
\end{center}}
\hspace*{0.2cm} \ddfermi{W^{(L)}}
%
%
\hspace*{0.2cm} + \hspace*{0.2cm}
\parbox{17mm}{\begin{center}
\begin{fmfgraph*}(14,8)
\setval \fmfstraight \fmfforce{2.5/14w,1.5/8h}{v1}
\fmfforce{2.5/14w,6.5/8h}{v2} \fmfforce{5/14w,1/2h}{v3}
\fmfforce{10/14w,1/2h}{v4} \fmfforce{1w,0h}{v5}
\fmfforce{1w,1h}{v6} \fmfforce{1w,1/2h}{v7}
\fmf{fermion,right=1}{v2,v1} \fmf{plain,right=1}{v1,v2}
\fmf{boson}{v4,v3} \fmf{fermion}{v5,v4} \fmf{fermion}{v4,v7}
\fmf{boson}{v4,v6} \fmfdot{v3,v4} \fmfv{decor.size=0,
label=${\scs \bar{1}}$, l.dist=1mm, l.angle=0}{v6} \fmfv{decor.size=0,
label=${\scs 2}$, l.dist=1mm, l.angle=0}{v5} \fmfv{decor.size=0,
label=${\scs 1}$, l.dist=1mm, l.angle=0}{v7}
\end{fmfgraph*}
\end{center}}
\hspace*{0.3cm}\dvertex{W^{(L)}}{$\bar{1}$}{2}{1}
%
%
\hspace*{0.2cm} + \hspace*{0.2cm}
\parbox{8mm}{\begin{center}
\begin{fmfgraph*}(5,13)
\setval \fmfstraight \fmfforce{0w,7/8h}{v1a}
\fmfforce{0w,1/4h}{v1b} \fmfforce{1w,0h}{v2}
\fmfforce{1w,1/4h}{v3} \fmfforce{1w,2/4h}{v4}
\fmfforce{1w,3/4h}{v5} \fmfforce{1w,1h}{v6} \fmf{fermion}{v2,v1b}
\fmf{fermion}{v1b,v3} \fmf{boson}{v4,v1b} \fmf{fermion}{v5,v1a}
\fmf{fermion}{v1a,v6} \fmf{boson}{v1a,v1b} \fmfdot{v1a,v1b}
\fmfv{decor.size=0, label=${\scs 1}$, l.dist=1mm, l.angle=0}{v6}
\fmfv{decor.size=0, label=${\scs 2}$, l.dist=1mm, l.angle=0}{v5}
\fmfv{decor.size=0, label=${\scs \bar{1}}$, l.dist=1mm, l.angle=0}{v4}
\fmfv{decor.size=0, label=${\scs 3}$, l.dist=1mm, l.angle=0}{v3}
\fmfv{decor.size=0, label=${\scs 4}$, l.dist=1mm, l.angle=0}{v2}
\end{fmfgraph*}
\end{center}}
\hspace*{0.3cm} \ddvertex{W^{(L)}}{1}{2}{$\bar{1}$}{3}{4}
\no \\
&& \hspace*{1cm}
%
%
+ \hspace*{0.1cm} \frac{1}{2}
\hspace*{0.1cm} \sum_{L'=2}^{L-1}
\hspace*{0.1cm}\dfermi{W^{(L')}}{1}{2} \hspace*{0.3cm}
\parbox{11mm}{\begin{center}
\begin{fmfgraph*}(8,8)
\setval \fmfstraight \fmfforce{0w,0h}{v1} \fmfforce{0w,1h}{v2}
\fmfforce{1/2w,1/2h}{v3} \fmfforce{1w,0h}{v4} \fmfforce{1w,1h}{v5}
\fmf{fermion}{v1,v3} \fmf{fermion}{v3,v2} \fmf{fermion}{v4,v3}
\fmf{fermion}{v3,v5} \fmfdot{v3} \fmfv{decor.size=0, label=${\scs
1}$, l.dist=1mm, l.angle=-180}{v2} \fmfv{decor.size=0,
label=${\scs 2}$, l.dist=1mm, l.angle=-180}{v1}
\fmfv{decor.size=0, label=${\scs 3}$, l.dist=1mm, l.angle=0}{v5}
\fmfv{decor.size=0, label=${\scs 4}$, l.dist=1mm, l.angle=0}{v4}
\end{fmfgraph*}
\end{center}}
\hspace*{0.3cm}\dfermi{W^{(L-L'+1)}}{3}{4}
%
%
\hspace*{0.2cm} + \hspace*{0.2cm}
\sum_{L'=2}^{L-1} \hspace*{0.1cm}\dfermi{W^{(L')}}{1}{2}
\hspace*{0.3cm}
\parbox{11mm}{\begin{center}
\begin{fmfgraph*}(8,8)
\setval \fmfstraight \fmfforce{0w,0h}{v1} \fmfforce{0w,1h}{v2}
\fmfforce{1/2w,1/2h}{v3} \fmfforce{1w,0h}{v4} \fmfforce{1w,1h}{v5}
\fmf{fermion}{v1,v3} \fmf{fermion}{v3,v2} \fmf{boson}{v4,v3}
\fmf{boson}{v3,v5} \fmfdot{v3} \fmfv{decor.size=0, label=${\scs
1}$, l.dist=1mm, l.angle=-180}{v2} \fmfv{decor.size=0,
label=${\scs 2}$, l.dist=1mm, l.angle=-180}{v1}
\fmfv{decor.size=0, label=${\scs \bar{1}}$, l.dist=1mm, l.angle=0}{v5}
\fmfv{decor.size=0, label=${\scs \bar{2}}$, l.dist=1mm, l.angle=0}{v4}
\end{fmfgraph*}
\end{center}}
\hspace*{0.3cm}\dbphi{W^{(L-L'+1)}}{$\bar{1}$}{$\bar{2}$}
\no \\ && \hspace*{1cm}
%
%
+  \hspace*{0.2cm} \sum_{L'=2}^{L-1}
\hspace*{0.1cm}\dfermi{W^{(L')}}{1}{2} \hspace*{0.3cm}
\parbox{16mm}{\begin{center}
\begin{fmfgraph*}(13,8)
\setval \fmfstraight \fmfforce{0w,0h}{v1} \fmfforce{0w,1h}{v2}
\fmfforce{4/13w,1/2h}{v3} \fmfforce{9/13w,1/2h}{v4}
\fmfforce{1w,0h}{v5} \fmfforce{1w,1h}{v6} \fmf{fermion}{v1,v3}
\fmf{fermion}{v3,v2} \fmf{boson}{v4,v3} \fmf{fermion}{v5,v4}
\fmf{fermion}{v4,v6} \fmfdot{v3,v4} \fmfv{decor.size=0,
label=${\scs 1}$, l.dist=1mm, l.angle=-180}{v2}
\fmfv{decor.size=0, label=${\scs 2}$, l.dist=1mm,
l.angle=-180}{v1} \fmfv{decor.size=0, label=${\scs 3}$,
l.dist=1mm, l.angle=0}{v6} \fmfv{decor.size=0, label=${\scs 4}$,
l.dist=1mm, l.angle=0}{v5}
\end{fmfgraph*}
\end{center}}
\hspace*{0.3cm}\dfermi{W^{(L-L'+1)}}{3}{4}
%
%
\hspace*{0.2cm} + \hspace*{0.2cm}
\sum_{L'=2}^{L-1} \hspace*{0.1cm}\dfermi{W^{(L')}}{1}{2}
\hspace*{0.3cm}
\parbox{16mm}{\begin{center}
\begin{fmfgraph*}(13,8)
\setval \fmfstraight \fmfforce{0w,0h}{v1} \fmfforce{0w,1h}{v2}
\fmfforce{4/13w,1/2h}{v3} \fmfforce{9/13w,1/2h}{v4}
\fmfforce{1w,0h}{v5} \fmfforce{1w,1h}{v6} \fmfforce{1w,1/2h}{v7}
\fmf{fermion}{v1,v3} \fmf{fermion}{v3,v2} \fmf{boson}{v4,v3}
\fmf{fermion}{v5,v4} \fmf{fermion}{v4,v7} \fmf{boson}{v4,v6}
\fmfdot{v3,v4} \fmfv{decor.size=0, label=${\scs 1}$, l.dist=1mm,
l.angle=-180}{v2} \fmfv{decor.size=0, label=${\scs 2}$,
l.dist=1mm, l.angle=-180}{v1} \fmfv{decor.size=0, label=${\scs
\bar{1}}$, l.dist=1mm, l.angle=0}{v6} \fmfv{decor.size=0, label=${\scs
4}$, l.dist=1mm, l.angle=0}{v5} \fmfv{decor.size=0, label=${\scs
3}$, l.dist=1mm, l.angle=0}{v7}
\end{fmfgraph*}
\end{center}}
\hspace*{0.3cm}\dvertex{W^{(L-L'+1)}}{$\bar{1}$}{4}{3} \Biggr],
\label{grapheq2}
\eeqn
\end{fmffile}
which we have written directly as a graphical equation where the
vertices already contain the coupling constants $\e,\esqr,g$. 
The last
four terms are the nonlinear ones. They start to give a
contribution only at the three-loop level. (Although the  overall
summation starts at $L'=2$, the nonlinear terms have an internal
summation for $L'\in [2,L-1]$, leading to a vanishing two-loop
contribution.) The fact that only positive signs enter is a
consequence of the Feynman rules (\ref{VR1})--(\ref{VR2}).

The term $\WL{2}$, corresponding to two-loop diagrams, is the
initial condition to enter this equation. It needs not to be given
because it is  also contained in~(\ref{grapheq2}). Identifying the 
terms of order $g$, we see directly that only the first line survives.
Integrating the equation gives the four diagrams in the right-hand-side, with
the corresponding weights $1/2,1/2,1/2,1/2$ obtained by dividing each
graph by its number of scalar lines (compare Table I):
\begin{fmffile}{graph10}
\begin{eqnarray}
W^{(2)} = 
%
%
\frac{1}{2}
\parbox{7mm}{\begin{center}
\begin{fmfgraph}(4,4)
\setval
\fmfforce{0w,1/2h}{v1}
\fmfforce{1w,1/2h}{v2}
\fmf{fermion,right=1}{v2,v1} 
\fmf{fermion,right=1}{v1,v2} 
\fmf{photon}{v2,v1}
\fmfdot{v2,v1}
\end{fmfgraph}\end{center}}
\hspace*{0.2cm} + \hspace*{0.2cm}
%
%
\frac{1}{2}
\parbox{15mm}{\begin{center}
\begin{fmfgraph}(12,4)
\setval
\fmfforce{1/6w,1h}{v1a}
\fmfforce{1/6w,0h}{v1b}
\fmfforce{1/3w,1/2h}{v2}
\fmfforce{2/3w,1/2h}{v3}
\fmfforce{5/6w,1h}{v4a}
\fmfforce{5/6w,0h}{v4b}
\fmf{fermion,right=1}{v1a,v1b} 
\fmf{plain,right=1}{v1b,v1a} 
\fmf{plain,right=1}{v1,v2}
\fmf{photon}{v2,v3}
\fmf{fermion,right=1}{v4b,v4a} 
\fmf{plain,right=1}{v4a,v4b}
\fmfdot{v2,v3}
\end{fmfgraph}\end{center}}
\hspace*{0.2cm} + \hspace*{0.2cm}
%
%
\frac{1}{2}\,
\parbox{11mm}{\begin{center}
\begin{fmfgraph}(8,4)
\setval
\fmfforce{1/4w,1h}{v1a}
\fmfforce{1/4w,0h}{v1b}
\fmfforce{1/2w,1/2h}{v2}
\fmfforce{1w,1/2h}{v3}
\fmf{fermion,right=1}{v1a,v1b} 
\fmf{plain,right=1}{v1b,v1a}
\fmf{photon,right=1}{v2,v3,v2}
\fmfdot{v2}
\end{fmfgraph}\end{center}}
\hspace*{0.2cm} + \hspace*{0.2cm}
%
%
\frac{1}{2}\,
\parbox{11mm}{\begin{center}
\begin{fmfgraph}(8,4)
\setval
\fmfforce{1/4w,1h}{v1a}
\fmfforce{1/4w,0h}{v1b}
\fmfforce{1/2w,1/2h}{v2}
\fmfforce{3/4w,1h}{v3a}
\fmfforce{3/4w,0h}{v3b}
\fmf{fermion,right=1}{v1a,v1b} 
\fmf{plain,right=1}{v1b,v1a}
\fmf{plain,right=1}{v3a,v3b}
\fmf{fermion,right=1}{v3b,v3a}
\fmfdot{v2}
\end{fmfgraph}\end{center}}
\end{eqnarray}
Equating powers of  $\g$, we
end up with the following nonlinear graphical recursion relation
for the vacuum diagrams:
\beqn
&&
\parbox{8mm}{\begin{center}
\begin{fmfgraph*}(2.5,5)
\setval \fmfstraight \fmfforce{1w,0h}{v1} \fmfforce{1w,1h}{v2}
\fmf{fermion,left=1}{v1,v2} \fmfv{decor.size=0, label=${\scs 2}$,
l.dist=1mm, l.angle=0}{v1} \fmfv{decor.size=0, label=${\scs 1}$,
l.dist=1mm, l.angle=0}{v2}
\end{fmfgraph*}
\end{center}}
\hspace*{0.3cm} \dfermi{W^{(L+1)}}{1}{2} \quad = \quad
%
%
2 \hspace*{0.2cm}
\parbox{12mm}{\begin{center}
\begin{fmfgraph*}(10,8)
\setval \fmfstraight \fmfforce{2.5/9w,1.5/8h}{v1}
\fmfforce{2.5/9w,6.5/8h}{v2} \fmfforce{5/9w,1/2h}{v3}
\fmfforce{1w,0h}{v4} \fmfforce{1w,1h}{v5}
\fmf{fermion,left=1}{v1,v2} \fmf{plain,left=1}{v2,v1}
\fmf{fermion}{v3,v5} \fmf{fermion}{v4,v3} \fmfdot{v3}
\fmfv{decor.size=0, label=${\scs 2}$, l.dist=1mm, l.angle=0}{v4}
\fmfv{decor.size=0, label=${\scs 1}$, l.dist=1mm, l.angle=0}{v5}
\end{fmfgraph*}
\end{center}}
\hspace*{0.3cm} \dfermi{W^{(L)}}{1}{2}
%
%
\hspace*{0.2cm} + \hspace*{0.1cm} \frac{1}{2} \hspace*{0.2cm}
\parbox{8mm}{\begin{center}
\begin{fmfgraph*}(5,10)
\setval \fmfstraight \fmfforce{0w,1/2h}{v1} \fmfforce{1w,0h}{v2}
\fmfforce{1w,1/3h}{v3} \fmfforce{1w,2/3h}{v4} \fmfforce{1w,1h}{v5}
\fmf{fermion}{v2,v1} \fmf{fermion}{v1,v3} \fmf{fermion}{v4,v1}
\fmf{fermion}{v1,v5} \fmfdot{v1} \fmfv{decor.size=0, label=${\scs
1}$, l.dist=1mm, l.angle=0}{v5} \fmfv{decor.size=0, label=${\scs
2}$, l.dist=1mm, l.angle=0}{v4} \fmfv{decor.size=0, label=${\scs
3}$, l.dist=1mm, l.angle=0}{v3} \fmfv{decor.size=0, label=${\scs
4}$, l.dist=1mm, l.angle=0}{v2}
\end{fmfgraph*}
\end{center}}
\hspace*{0.2cm} \ddfermi{W^{(L)}}
%
%
\hspace*{0.2cm} + \hspace*{0.1cm} \frac{1}{2} \hspace*{0.2cm}
\parbox{12mm}{\begin{center}
\begin{fmfgraph*}(10,8)
\setval \fmfstraight \fmfforce{2.5/9w,1.5/8h}{v1}
\fmfforce{2.5/9w,6.5/8h}{v2} \fmfforce{5/9w,1/2h}{v3}
\fmfforce{1w,0h}{v4} \fmfforce{1w,1h}{v5}
\fmf{boson,left=1}{v1,v2,v1} \fmf{fermion}{v3,v5}
\fmf{fermion}{v4,v3} \fmfdot{v3} \fmfv{decor.size=0, label=${\scs
2}$, l.dist=1mm, l.angle=0}{v4} \fmfv{decor.size=0, label=${\scs
1}$, l.dist=1mm, l.angle=0}{v5}
\end{fmfgraph*}
\end{center}}
\hspace*{0.3cm} \dfermi{W^{(L)}}{1}{2}
\no \\ && \hspace*{1cm}
%
%
+ \hspace*{0.2cm}
\parbox{12mm}{\begin{center}
\begin{fmfgraph*}(9,8)
\setval \fmfstraight \fmfforce{2.5/9w,1.5/8h}{v1}
\fmfforce{2.5/9w,6.5/8h}{v2} \fmfforce{5/9w,1/2h}{v3}
\fmfforce{1w,0h}{v4} \fmfforce{1w,1h}{v5}
\fmf{fermion,left=1}{v1,v2} \fmf{plain,left=1}{v2,v1}
\fmf{boson}{v3,v5} \fmf{boson}{v4,v3} \fmfdot{v3}
\fmfv{decor.size=0, label=${\scs \bar{2}}$, l.dist=1mm, l.angle=0}{v4}
\fmfv{decor.size=0, label=${\scs \bar{1}}$, l.dist=1mm, l.angle=0}{v5}
\end{fmfgraph*}
\end{center}}
\hspace*{0.3cm} \dbphi{W^{(L)}}{$\bar{1}$}{$\bar{2}$}
%
%
\hspace*{0.2cm} + \hspace*{0.2cm}
\parbox{12mm}{\begin{center}
\begin{fmfgraph*}(5,10)
\setval \fmfstraight \fmfforce{0w,1/2h}{v1} \fmfforce{1w,0h}{v2}
\fmfforce{1w,1/3h}{v3} \fmfforce{1w,2/3h}{v4} \fmfforce{1w,1h}{v5}
\fmf{boson}{v2,v1} \fmf{boson}{v1,v3} \fmf{fermion}{v4,v1}
\fmf{fermion}{v1,v5} \fmfdot{v1} \fmfv{decor.size=0, label=${\scs
1}$, l.dist=1mm, l.angle=0}{v5} \fmfv{decor.size=0, label=${\scs
2}$, l.dist=1mm, l.angle=0}{v4} \fmfv{decor.size=0, label=${\scs
\bar{1}}$, l.dist=1mm, l.angle=0}{v3} \fmfv{decor.size=0, label=${\scs
\bar{2}}$, l.dist=1mm, l.angle=0}{v2}
\end{fmfgraph*}
\end{center}}
\hspace*{0.2cm} \ddbfermi{W^{(L)}}
%
%
\hspace*{0.2cm} + \hspace*{0.1cm} 2 \hspace*{0.2cm}
\parbox{17mm}{\begin{center}
\begin{fmfgraph*}(14,8)
\setval \fmfstraight \fmfforce{2.5/14w,1.5/8h}{v1}
\fmfforce{2.5/14w,6.5/8h}{v2} \fmfforce{5/14w,1/2h}{v3}
\fmfforce{10/14w,1/2h}{v4} \fmfforce{1w,0h}{v5}
\fmfforce{1w,1h}{v6} \fmf{fermion,right=1}{v2,v1}
\fmf{plain,left=1}{v2,v1} \fmf{boson}{v3,v4} \fmf{fermion}{v5,v4}
\fmf{fermion}{v4,v6} \fmfdot{v3,v4} \fmfv{decor.size=0,
label=${\scs 1}$, l.dist=1mm, l.angle=0}{v6} \fmfv{decor.size=0,
label=${\scs 2}$, l.dist=1mm, l.angle=0}{v5}
\end{fmfgraph*}
\end{center}}
\hspace*{0.2cm} \dfermi{W^{(L)}}{1}{2}
%
%
\hspace*{0.2cm} + \hspace*{0.1cm} 2 \hspace*{0.2cm}
\parbox{10.5mm}{\begin{center}
\begin{fmfgraph*}(7.5,5)
\setval \fmfstraight \fmfforce{1/3w,0h}{v1} \fmfforce{1/3w,1h}{v2}
\fmfforce{1w,0h}{v3} \fmfforce{1w,1h}{v4}
\fmf{fermion,left=1}{v1,v2} \fmf{fermion}{v2,v4}
\fmf{fermion}{v3,v1} \fmf{boson}{v1,v2} \fmfdot{v1,v2}
\fmfv{decor.size=0, label=${\scs 1}$, l.dist=1mm, l.angle=0}{v4}
\fmfv{decor.size=0, label=${\scs 2}$, l.dist=1mm, l.angle=0}{v3}
\end{fmfgraph*}
\end{center}}
\hspace*{0.2cm} \dfermi{W^{(L)}}{1}{2}
\no \\ && \hspace*{1cm}
%
%
+ \hspace*{0.2cm}
\parbox{8mm}{\begin{center}
\begin{fmfgraph*}(5,10)
\setval \fmfstraight \fmfforce{0w,1/6h}{v1a}
\fmfforce{0w,5/6h}{v1b} \fmfforce{1w,0h}{v2}
\fmfforce{1w,1/3h}{v3} \fmfforce{1w,2/3h}{v4} \fmfforce{1w,1h}{v5}
\fmf{fermion}{v2,v1a} \fmf{fermion}{v1a,v3} \fmf{fermion}{v4,v1b}
\fmf{fermion}{v1b,v5} \fmf{boson}{v1a,v1b} \fmfdot{v1a,v1b}
\fmfv{decor.size=0, label=${\scs 1}$, l.dist=1mm, l.angle=0}{v5}
\fmfv{decor.size=0, label=${\scs 2}$, l.dist=1mm, l.angle=0}{v4}
\fmfv{decor.size=0, label=${\scs 3}$, l.dist=1mm, l.angle=0}{v3}
\fmfv{decor.size=0, label=${\scs 4}$, l.dist=1mm, l.angle=0}{v2}
\end{fmfgraph*}
\end{center}}
\hspace*{0.2cm} \ddfermi{W^{(L)}}
%
%
\hspace*{0.2cm} + \hspace*{0.2cm}
\parbox{17mm}{\begin{center}
\begin{fmfgraph*}(14,8)
\setval \fmfstraight \fmfforce{2.5/14w,1.5/8h}{v1}
\fmfforce{2.5/14w,6.5/8h}{v2} \fmfforce{5/14w,1/2h}{v3}
\fmfforce{10/14w,1/2h}{v4} \fmfforce{1w,0h}{v5}
\fmfforce{1w,1h}{v6} \fmfforce{1w,1/2h}{v7}
\fmf{fermion,right=1}{v2,v1} \fmf{plain,right=1}{v1,v2}
\fmf{boson}{v4,v3} \fmf{fermion}{v5,v4} \fmf{fermion}{v4,v7}
\fmf{boson}{v4,v6} \fmfdot{v3,v4} \fmfv{decor.size=0,
label=${\scs \bar{1}}$, l.dist=1mm, l.angle=0}{v6} \fmfv{decor.size=0,
label=${\scs 2}$, l.dist=1mm, l.angle=0}{v5} \fmfv{decor.size=0,
label=${\scs 1}$, l.dist=1mm, l.angle=0}{v7}
\end{fmfgraph*}
\end{center}}
\hspace*{0.3cm}\dvertex{W^{(L)}}{$\bar{1}$}{2}{1}
%
%
\hspace*{0.2cm} + \hspace*{0.2cm}
\parbox{8mm}{\begin{center}
\begin{fmfgraph*}(5,13)
\setval \fmfstraight \fmfforce{0w,7/8h}{v1a}
\fmfforce{0w,1/4h}{v1b} \fmfforce{1w,0h}{v2}
\fmfforce{1w,1/4h}{v3} \fmfforce{1w,2/4h}{v4}
\fmfforce{1w,3/4h}{v5} \fmfforce{1w,1h}{v6} \fmf{fermion}{v2,v1b}
\fmf{fermion}{v1b,v3} \fmf{boson}{v4,v1b} \fmf{fermion}{v5,v1a}
\fmf{fermion}{v1a,v6} \fmf{boson}{v1a,v1b} \fmfdot{v1a,v1b}
\fmfv{decor.size=0, label=${\scs 1}$, l.dist=1mm, l.angle=0}{v6}
\fmfv{decor.size=0, label=${\scs 2}$, l.dist=1mm, l.angle=0}{v5}
\fmfv{decor.size=0, label=${\scs \bar{1}}$, l.dist=1mm, l.angle=0}{v4}
\fmfv{decor.size=0, label=${\scs 3}$, l.dist=1mm, l.angle=0}{v3}
\fmfv{decor.size=0, label=${\scs 4}$, l.dist=1mm, l.angle=0}{v2}
\end{fmfgraph*}
\end{center}}
\hspace*{0.3cm} \ddvertex{W^{(L)}}{1}{2}{$\bar{1}$}{3}{4} \no \\ &&
\hspace*{1cm}
%
%
+ \hspace*{0.1cm} \frac{1}{2} \hspace*{0.1cm} \sum_{L'=2}^{L-1}
\hspace*{0.1cm}\dfermi{W^{(L')}}{1}{2} \hspace*{0.3cm}
\parbox{11mm}{\begin{center}
\begin{fmfgraph*}(8,8)
\setval \fmfstraight \fmfforce{0w,0h}{v1} \fmfforce{0w,1h}{v2}
\fmfforce{1/2w,1/2h}{v3} \fmfforce{1w,0h}{v4} \fmfforce{1w,1h}{v5}
\fmf{fermion}{v1,v3} \fmf{fermion}{v3,v2} \fmf{fermion}{v4,v3}
\fmf{fermion}{v3,v5} \fmfdot{v3} \fmfv{decor.size=0, label=${\scs
1}$, l.dist=1mm, l.angle=-180}{v2} \fmfv{decor.size=0,
label=${\scs 2}$, l.dist=1mm, l.angle=-180}{v1}
\fmfv{decor.size=0, label=${\scs 3}$, l.dist=1mm, l.angle=0}{v5}
\fmfv{decor.size=0, label=${\scs 4}$, l.dist=1mm, l.angle=0}{v4}
\end{fmfgraph*}
\end{center}}
\hspace*{0.3cm}\dfermi{W^{(L-L'+1)}}{3}{4}
%
%
\hspace*{0.2cm} + \hspace*{0.2cm}
\sum_{L'=2}^{L-1} \hspace*{0.1cm}\dfermi{W^{(L')}}{1}{2}
\hspace*{0.3cm}
\parbox{11mm}{\begin{center}
\begin{fmfgraph*}(8,8)
\setval \fmfstraight \fmfforce{0w,0h}{v1} \fmfforce{0w,1h}{v2}
\fmfforce{1/2w,1/2h}{v3} \fmfforce{1w,0h}{v4} \fmfforce{1w,1h}{v5}
\fmf{fermion}{v1,v3} \fmf{fermion}{v3,v2} \fmf{boson}{v4,v3}
\fmf{boson}{v3,v5} \fmfdot{v3} \fmfv{decor.size=0, label=${\scs
1}$, l.dist=1mm, l.angle=-180}{v2} \fmfv{decor.size=0,
label=${\scs 2}$, l.dist=1mm, l.angle=-180}{v1}
\fmfv{decor.size=0, label=${\scs \bar{1}}$, l.dist=1mm, l.angle=0}{v5}
\fmfv{decor.size=0, label=${\scs \bar{2}}$, l.dist=1mm, l.angle=0}{v4}
\end{fmfgraph*}
\end{center}}
\hspace*{0.3cm}\dbphi{W^{(L-L'+1)}}{$\bar{1}$}{$\bar{2}$}
\no \\ && \hspace*{1cm}
%
%
+ \hspace*{0.2cm}  \sum_{L'=2}^{L-1}
\hspace*{0.1cm}\dfermi{W^{(L')}}{1}{2} \hspace*{0.3cm}
\parbox{16mm}{\begin{center}
\begin{fmfgraph*}(13,8)
\setval \fmfstraight \fmfforce{0w,0h}{v1} \fmfforce{0w,1h}{v2}
\fmfforce{4/13w,1/2h}{v3} \fmfforce{9/13w,1/2h}{v4}
\fmfforce{1w,0h}{v5} \fmfforce{1w,1h}{v6} \fmf{fermion}{v1,v3}
\fmf{fermion}{v3,v2} \fmf{boson}{v4,v3} \fmf{fermion}{v5,v4}
\fmf{fermion}{v4,v6} \fmfdot{v3,v4} \fmfv{decor.size=0,
label=${\scs 1}$, l.dist=1mm, l.angle=-180}{v2}
\fmfv{decor.size=0, label=${\scs 2}$, l.dist=1mm,
l.angle=-180}{v1} \fmfv{decor.size=0, label=${\scs 3}$,
l.dist=1mm, l.angle=0}{v6} \fmfv{decor.size=0, label=${\scs 4}$,
l.dist=1mm, l.angle=0}{v5}
\end{fmfgraph*}
\end{center}}
\hspace*{0.3cm}\dfermi{W^{(L-L'+1)}}{3}{4}
%
%
\hspace*{0.2cm} + \hspace*{0.2cm}
\sum_{L'=2}^{L-1} \hspace*{0.1cm}\dfermi{W^{(L')}}{1}{2}
\hspace*{0.3cm}
\parbox{16mm}{\begin{center}
\begin{fmfgraph*}(13,8)
\setval \fmfstraight \fmfforce{0w,0h}{v1} \fmfforce{0w,1h}{v2}
\fmfforce{4/13w,1/2h}{v3} \fmfforce{9/13w,1/2h}{v4}
\fmfforce{1w,0h}{v5} \fmfforce{1w,1h}{v6} \fmfforce{1w,1/2h}{v7}
\fmf{fermion}{v1,v3} \fmf{fermion}{v3,v2} \fmf{boson}{v4,v3}
\fmf{fermion}{v5,v4} \fmf{fermion}{v4,v7} \fmf{boson}{v4,v6}
\fmfdot{v3,v4} \fmfv{decor.size=0, label=${\scs 1}$, l.dist=1mm,
l.angle=-180}{v2} \fmfv{decor.size=0, label=${\scs 2}$,
l.dist=1mm, l.angle=-180}{v1} \fmfv{decor.size=0, label=${\scs
\bar{1}}$, l.dist=1mm, l.angle=0}{v6} \fmfv{decor.size=0, label=${\scs
4}$, l.dist=1mm, l.angle=0}{v5} \fmfv{decor.size=0, label=${\scs
3}$, l.dist=1mm, l.angle=0}{v7}
\end{fmfgraph*}
\end{center}}
\hspace*{0.3cm}\dvertex{W^{(L-L'+1)}}{$\bar{1}$}{4}{3}.
\label{grapheq5}
\eeqn
\end{fmffile}

The recursive nature of this equation is due to the fact that the
left-hand-side  is of one higher loop order than the right-hand-side:
To obtain the diagrams at order $L+1$, one applies the  right-hand-side 
operators on the
lower order diagrams. Since the left-hand-side contains the 
scalar number operator 
for each diagram, the weight of the diagram which have just been
obtained has to be divided by the respective number of scalar lines, see
Eq.~(\ref{countingquarklines}). We remind the reader that this 
number can be obtained knowing
the number of seagull vertices of the respective diagram, see Eq.~(\ref{Eqquarknumber}).

We have derived the vacuum diagrams of the \gl
theory up to four loops which involves the nonlinear
part of Eq.~(\ref{grapheq5}). Indeed, the nonlinear terms only
enter from the three-loop order. The resulting
diagrams are given in Table~1. The effort needed to obtain them is
considerably reduced compared to a calculation done with the help of external sources coupled linearily to the fields.
Note that Eq.~(\ref{grapheq5}) is
suitable for an automatized symbolic computation which can be implemented as in Ref. \cite{phi4}, 
such that one may proceed to higher orders without much
effort except for computer time. 

A look at Table 1 shows that the diagrams and weights
of the Yukawa part coincide with those from Ref.~\cite{qed}, as it
should. For the pure $\phi^4$-part, the diagrams are equivalent
to those in Ref.~\cite{phi4}, although the weights do not
coincide, since we deal with complex scalar fields.

Being in the possession of all Feynman diagrams we must still
calculate the associated integrals in order to extract physical results.
This will be done in a separate publication. In particular, we intend to compute the vacuum energy which
determines the critical behavior of the heat capacity of a superconductor at the phase transition.

\section{Conclusions}
\label{conclusions}

In this paper, we set up a graphical recursion relation for obtaining the
connected diagrams of the \gl model which describes superconductors near the critical point. We have used our
equation to obtain the diagrams up to the four-loop order.
These diagrams will be needed to extend our two-loop calculations in
Ref.~\cite{kvdbeffpot} to higher orders.

\section*{Acknowledgement}

The work of B.V.d.B. was supported by the Alexander von Humboldt foundation and the Institut Interuniversitaire
des Sciences Nucl\'eaires de Belgique.

\newpage

\begin{table}[t]
\begin{center}
\begin{tabular}{|cccc|c|}
\,\,\,$L$ &
$n_1$ &
$n_2$ &
$n_3$ &
$W^{(L,n_1,n_2,n_3)}$
\\
\hline
$2$ & $0$ & $0$ & $2$ &
\begin{fmffile}{gl2002a}
\hspace{-10pt}
\rule[-10pt]{0pt}{26pt}
%
%
${\displaystyle \frac{1}{2}}$
\parbox{7mm}{\begin{center}
\begin{fmfgraph}(4,4)
\setval
\fmfforce{0w,1/2h}{v1}
\fmfforce{1w,1/2h}{v2}
\fmf{fermion,right=1}{v2,v1} 
\fmf{fermion,right=1}{v1,v2} 
\fmf{photon}{v2,v1}
\fmfdot{v2,v1}
\end{fmfgraph}\end{center}}
%
%
${\displaystyle \frac{1}{2}}$
\parbox{15mm}{\begin{center}
\begin{fmfgraph}(12,4)
\setval
\fmfforce{1/6w,1h}{v1a}
\fmfforce{1/6w,0h}{v1b}
\fmfforce{1/3w,1/2h}{v2}
\fmfforce{2/3w,1/2h}{v3}
\fmfforce{5/6w,1h}{v4a}
\fmfforce{5/6w,0h}{v4b}
\fmf{fermion,right=1}{v1a,v1b} 
\fmf{plain,right=1}{v1b,v1a} 
\fmf{plain,right=1}{v1,v2}
\fmf{photon}{v2,v3}
\fmf{fermion,right=1}{v4b,v4a} 
\fmf{plain,right=1}{v4a,v4b}
\fmfdot{v2,v3}
\end{fmfgraph}\end{center}}
\end{fmffile}
\\ 
$2$ & $0$ & $1$ & $0$ &
\hspace{-10pt}
\rule[-10pt]{0pt}{26pt}
\begin{fmffile}{gl2010a}
${\displaystyle \frac{1}{2}}$
\parbox{11mm}{\begin{center}
\begin{fmfgraph}(8,4)
\setval
\fmfforce{1/4w,1h}{v1a}
\fmfforce{1/4w,0h}{v1b}
\fmfforce{1/2w,1/2h}{v2}
\fmfforce{1w,1/2h}{v3}
\fmf{fermion,right=1}{v1a,v1b} 
\fmf{plain,right=1}{v1b,v1a}
\fmf{photon,right=1}{v2,v3,v2}
\fmfdot{v2}
\end{fmfgraph}\end{center}}
\end{fmffile}
\\ 
$2$ & $1$ & $0$ & $0$ & 
\hspace{-10pt}
\rule[-10pt]{0pt}{26pt}
\begin{fmffile}{gl2100a}
${\displaystyle \frac{1}{2}}$\,
\parbox{11mm}{\begin{center}
\begin{fmfgraph}(8,4)
\setval
\fmfforce{1/4w,1h}{v1a}
\fmfforce{1/4w,0h}{v1b}
\fmfforce{1/2w,1/2h}{v2}
\fmfforce{3/4w,1h}{v3a}
\fmfforce{3/4w,0h}{v3b}
\fmf{fermion,right=1}{v1a,v1b} 
\fmf{plain,right=1}{v1b,v1a}
\fmf{plain,right=1}{v3a,v3b}
\fmf{fermion,right=1}{v3b,v3a}
\fmfdot{v2}
\end{fmfgraph}\end{center}}
\end{fmffile}
\\
\end{tabular}
\end{center}
\caption{Connected vacuum diagrams $W^{(L,n_1,n_2,n_3)}$ and their weights up to the four-loop order
of the $O(2)$ \gl model, wehre $L$ denotes the loop order and $n_1,n_2,n_3$ count the number of vertices
$V,F,H$, respectively.}
\end{table}

\end{document}